\documentclass[12pt]{article}
\usepackage{hyperref}
\usepackage{graphicx}
\usepackage{color}
\usepackage{amsmath,amssymb} 

\definecolor{myblue}{rgb}{0.14,0.11,0.49}
\definecolor{myred}{rgb}{0.74,0.22,0.15}
\definecolor{mygreen}{rgb}{0.05,0.52,0.42}
\definecolor{myyellow}{rgb}{0.96,0.92,0.13}
\definecolor{myorange}{rgb}{1,0.61,0.36}
\definecolor{mypurple}{rgb}{0.71,0.02,1}
\definecolor{noir}{gray}{0.} 

\newcommand{\Couleur}[1]{\textcolor{noir}{#1}}

\definecolor{htc}{rgb}{1,1,1} 

\newcommand{\Mat}[1]{{{\boldsymbol{#1}}}}

\def\be{\begin{equation}}
\def\ee{\end{equation}}
\def\bea{\begin{equationarray}}
\def\eea{\end{equationarray}}
\def\bc{\begin{center}}
\def\ec{\end{center}}
\def\bi{\begin{itemize}}
\def\ei{\end{itemize}}
\def\bs{\begin{slide}}
\def\es{\end{slide}}
\def\dd{\mathrm{d}}
\def\iC{\mathrm{i}}
\def\noi{\noindent}
\date{}
\begin{document}
\title{On the Hamiltonian and energy operators in a curved spacetime, especially for a Dirac particle}
\author{Mayeul Arminjon\\
\small\it Laboratory ``Soils, Solids, Structures, Risks'', 3SR,\\ \small\it CNRS and Grenoble-Alpes University,\\ \small\it BP 53, F-38041 Grenoble cedex 9, France.}

\maketitle

\begin{abstract}
\noindent The definition of the Hamiltonian operator $\,\mathrm{H}\,$ for a general wave equation in a general spacetime is discussed. We recall that $\,\mathrm{H}\,$ depends on the coordinate system merely through the corresponding reference frame. When the wave equation involves a gauge choice and the gauge change is time-dependent, $\,\mathrm{H}\,$ as an operator depends on the gauge choice. This dependence extends to the energy operator $\,\mathrm{E}$, which is the Hermitian part of $\,\mathrm{H}$. We distinguish between this ambiguity issue of $\,\mathrm{E}\,$ and the one that occurs due to a mere change of the ``representation" (e.g. transforming the Dirac wave function from the ``Dirac representation" to a ``Foldy-Wouthuysen representation"). We also assert that the energy operator ought to be well defined in a given reference frame at a given time, e.g. by comparing the situation for this operator with the main features of the energy for a classical Hamiltonian particle. 

\end{abstract}

\section{Introduction}
          
\noindent The quantum effects in the classical gravitational field which have been observed on Earth for neutrons and for atoms (e.g. \cite{COW1975, WernerStaudenmannColella1979, RiehleBorde1991, KasevichChu1991, Nesvizhevsky2002}) are quantum-mechanical effects. That is, they were predicted (before their observation) by using ``first-quantized" theory. They are, to this author's knowledge, the only observed effects of the gravity-quantum coupling. Thus, QM in a curved spacetime covers all currently available experiments about the interaction between gravity and the quantum. Neutrons are spin half particles, and spin half particles are normally described by the Dirac equation. Moreover, besides observing directly the quantum behaviour of particles in a gravitational field, one also would like to account for the influence of the gravitational field on, for example, the quantum behaviour of particles in an electromagnetic field, e.g. the behaviour of electrons in an atom. (This influence is very small in the very weak gravitational field that we have in the solar system, of course. Yet it is expected to be important in strong gravitational fields.) Therefore, QM of the (generally-)covariant Dirac equation is quite an important chapter of curved-spacetime QM. The covariant Dirac equation involves the choice, at any point $X$ in spacetime (and depending smoothly on $X$), of an orthonormal basis of the tangent space at that point $X$. I.e., it involves the choice of a {\it tetrad field} \cite{BrillWheeler1957+Corr, ChapmanLeiter1976}. This plays the role of a gauge field. Indeed, the two realizations of the covariant Dirac equation that are got with two different tetrad fields are equivalent, at least locally \cite{Fock1929}. However, it has been proved in recent years that the Hamiltonian operator associated, in a given coordinate system, with the covariant Dirac equation, does depend on the choice of the tetrad field. In particular, the energy spectrum and the energy mean values depend on that choice \cite{A43,A50}, and so does the presence or absence of Mashhoon's spin-rotation coupling term \cite{Ryder2008,A49}.

In the present conference paper, we summarize the main part of the recent results \cite{A51} related with the non-uniqueness problem of the Hamiltonian and energy operators. We add some new remarks, especially we add a new discussion in Sect. \ref{Ambiguity E}. In Sect. \ref{Def H}, the definition of the Hamiltonian operator $\,\mathrm{H}\,$ for a general wave equation in a general spacetime is discussed. We recall that $\,\mathrm{H}\,$ depends on the coordinate system merely through the corresponding reference frame. Then in Sect. \ref{H with gauge} we consider the situation in which the wave equation involves a gauge choice (as is the case for the covariant Dirac equation). We show that, when the gauge transformation depends on the time coordinate, $\,\mathrm{H}\,$ as an operator depends on the gauge choice. This extends the previous results \cite{A43,A50} from the covariant Dirac equation to any quantum wave equation admitting a gauge choice. Section \ref{Ambiguity E} introduces the energy operator and discusses its ambiguity in the presence of a gauge choice. We distinguish between this ambiguity issue and the one that occurs due to a mere change of the ``representation" --- e.g. from the ``Dirac representation" to a ``Foldy-Wouthuysen representation". In Sect. \ref{Energy}, we assert that the energy operator ought to be well defined. In particular, recalling that in QM the meaning of the energy is inherited from classical Hamiltonian mechanics, we note that the classical Hamiltonian energy is in general not conserved and depends exactly on the reference frame as we define it. The same is true for the quantum-mechanical energy as it arises from the energy operator $\mathrm{E}$, as long as there is no gauge choice.

  \section{Hamiltonian operator in a general spacetime $\Couleur{\mathrm{V}}$}\label{Def H}

Consider a wave equation for a quantum particle in the spacetime manifold $\Couleur{\mathrm{V}}$. In order to define a Hamiltonian operator, one needs first to choose a local chart (coordinate system). This is in general defined only in some subdomain $\mathrm{U}$ of $\mathrm{V}$. To any point $X$ in $ \mathrm{U}$, the chart $\chi $ associates a quadruplet of real numbers: $\chi (X)={\bf X}\equiv (x^\mu ) \in \mathbb{R}^4$. In any such chart $\chi $, the wave operator gets a local expression, i.e., an expression as an usual linear differential operator acting on local wave functions $\,\Couleur{\Psi=\Psi ({\bf X})}$ defined on the open subset $ \chi (\mathrm{U})$ of $\mathbb{R}^4$ and taking values in some $\mathbb{C}^n$ space. 
\footnote{
We may assume that the ``intrinsic" wave function, say $\varphi $, takes value directly in this $\mathbb{C}^n$ space: Otherwise, if $\varphi $ is a section of some vector bundle with base $\mathrm{V}$, the components $\Psi ^a\ (a=1,...,n)$ of the local expression of $\varphi $ are got from the expansion of $\varphi$  on some local frame field $(e_a)$.  
}
Assume this local expression contains only first-order derivatives with respect to $\,\Couleur{t\equiv x^0/c}$. In that case, we may rewrite the wave equation as the Schr\"odinger equation (setting $\hbar =1$ in this paper): 
\be \label{Schrodinger-general}
\mathrm{i} \frac{\partial \Psi }{\partial t}= \mathrm{H}\Psi, 
\ee
in which the Hamiltonian operator $\,\mathrm{H}\,$ is a linear differential operator that contains no derivative with respect to $t$. This definition makes it clear that $\,\mathrm{H}\,$ depends on the coordinate system as do the local expressions of the wave function and of the wave operator.
Also, $\Couleur{\mathrm{H}}\,$ is defined only over the domain of the chart, $\,\Couleur{\mathrm{U}\subset \mathrm{V}}$. But in practice, $\,\Couleur{\mathrm{U}}$ can be taken large enough that we may forget about the values of $\,\Couleur{\Psi}\,$ outside $\,\Couleur{\mathrm{U}}$ (or rather, outside $\,\chi(\mathrm{U})\,$). More exactly, $\Psi (t,{\bf x})$ may be assumed to vanish when the triplet ${\bf x}$ made with the spatial coordinates $x^j$ ($j=1,2,3$) is on the spatial boundary of $\,\chi(\mathrm{U})\,$. (See Ref. \cite{A51} for more detail.)  If one changes the chart: $\Couleur{{\bf X}=(x^\mu )\hookrightarrow (x'^\mu ) ={\bf X}'}$, one gets another operator $\,\Couleur{\mathrm{H}' \ne \mathrm{H}}$, whose relation with $\,\mathrm{H}\,$ is not simple in general \cite{A42}. However, $\,\Couleur{\mathrm{H}}\,$ is essentially unchanged under a {\it purely spatial} coordinate change: 
\be\label{purely-spatial-change}
\Couleur{x'^j=f^j((x^k))\ (j,k=1,2,3)},\qquad \mathrm{and}\qquad \Couleur{x'^0=x^0}.
\ee
E.g., when $\,\Couleur{\Psi }\,$ behaves as a scalar, $\,\Couleur{\Psi'({\bf X}')=\Psi({\bf X})}$, we have:
\be\label{H'=H}
\Couleur{(\mathrm{H}'\Psi')({\bf X}')=(\mathrm{H}\Psi)({\bf X})}. 
\ee
We call ``reference frame" an {\it equivalence class $\,\Couleur{\mathrm{F}}\,$ of charts} defined on the same domain $\,\Couleur{\mathrm{U}}\,$ and exchanging by \,(\ref{purely-spatial-change}) \cite{A44}. A reference frame $\,\Couleur{\mathrm{F}}\,$ in that sense amounts to the data of:
\bi
\item the domain $\,\Couleur{\mathrm{U}}\,$ of the spacetime,

\item The set of the reference world lines ``$\Couleur{{\bf x}\equiv (x^j)=}\mathrm{\, Constant},\Couleur{\ x^0}\ \mathrm{variable}$", 

\item A time coordinate map $\,\Couleur{X \mapsto x^0(X)}\,$ defined for events $\,\Couleur{X\in \mathrm{U}}$.
\ei
So this extends the notion of reference frame of classical mechanics \cite{A51}.

\noi {\it The operator $\,\Couleur{\mathrm{H}}\,$ depends precisely on the reference frame in this sense} \cite{A42}.

  \section{Hamiltonians with a gauge choice}\label{H with gauge}

In this section and in the next one, we fix the reference frame, but we suppose from now that the wave operator depends on the choice of some ``gauge" field $\,\Couleur{G}$. This happens e.g. for the covariant Dirac equation, for which $\,\Couleur{G}\,$ is the tetrad field. We assume \cite{A51} that:
\bi

\item (i) For any choice of the gauge field $\,\Couleur{G}$, there is a Hilbert space $\,\Couleur{{\sf H}}\,$ of ``states" $\,\Couleur{\Psi =\Psi ({\bf x})}$, with scalar product: $\, \Couleur{(\Psi \mid \Phi )}$. Thus for another choice $\,\Couleur{\widetilde{G}}\,$ of the gauge field, we get another Hilbert space $\,\widetilde{{\sf H}}$, and another scalar product $\,(\Xi \, \widetilde{\mid}\, \Omega  )$. 
\footnote{
In general, the definition of the scalar product involves some coefficient {\it fields}, e.g. the field of the curved-spacetime Dirac matrix $\gamma ^0$ in the case of the covariant Dirac equation: for instance, see Eq. (26)$_1$ in Ref. \cite{A51}. It follows that in general, strictly speaking, the scalar product and even the Hilbert space $\Couleur{{\sf H}}$ depend on the value of the time coordinate $t$. We will keep this dependence implicit.
}

\vspace{1mm}
\item (ii) Given any two gauge fields, there is (for any value of $t$) a {\it unitary transformation} $\,\mathcal{U}\,$ from $\,{\sf H}\,$ onto $\,\widetilde{{\sf H}}\,$:
\be \label{tilde=isometry}
\forall t,\ \forall \Psi ,\Phi \in {\sf H},\quad (\mathcal{U}\Psi \,\widetilde { \mid}\,\mathcal{U}\Phi )=(\Psi \mid \Phi ).
\ee

\item (iii) The wave equation is covariant under any unitary transformation $\,\Couleur{\mathcal{U}}\,$ arising due to an admissible change of the gauge field: $\,\Couleur{G \hookrightarrow \widetilde{G}}$.

\ei
These assumptions are indeed valid in the case of the covariant Dirac equation \cite{A50}. When changing the gauge field: $\,\Couleur{G \hookrightarrow \widetilde{G}}$, we get a new form of the wave equation. Thus, when rewriting it as the Schr\"odinger equation, we get a new Hamiltonian operator: $\,\Couleur{\mathrm{H} \hookrightarrow \widetilde{\mathrm{H}}}$. The covariance of the wave equation under the corresponding unitary transformation $\,\Couleur{\mathcal{U}}\,$ means this:
\be\label{Psi or Xi}
\mathrm{If}\ \,\Couleur{\ \Xi \equiv \mathcal{U}\Psi},\quad \mathrm{then}\quad \Couleur{\mathrm{i} \frac{\partial \Psi }{\partial t}= \mathrm{H}\Psi\  \Leftrightarrow \  \mathrm{i} \frac{\partial \Xi  }{\partial t}= \widetilde{\mathrm{H}}\Xi\ }.
\ee
In turn, it is easy to check that (\ref{Psi or Xi}) is true if and only if:
\be\label{Htilde}
\underline{\Couleur{\widetilde{\mathrm{H}}=\mathcal{U} \mathrm{H}\mathcal{U}^{-1}-\mathrm{i}\,\mathcal{U}\left[\partial_t \left(\mathcal{U}^{-1}\right)\right]}}. 
\ee
Equation (\ref{Htilde}) is well known (in a somewhat different context, see Section \ref{Ambiguity E}) \cite{FoldyWouthuysen1950,Goldman1977}. It means that $\,\Couleur{\mathrm{H}}$, seen as the {\it generator of the time evolution,} transforms consistently. 

But, seen as an {\it operator acting on the states $\,\Couleur{\Psi =\Psi ({\bf x})}$,} $\,\Couleur{\mathrm{H}}\,$ would transform consistently if and only if, at any time $\,\Couleur{t}$, all mean values were invariant under the unitary transformation. Thus, $\,\Couleur{\mathcal{D}}\,$ being the domain of $\,\Couleur{\mathrm{H}}$, if and only if:
\be \label{H tilde equivt to H - mean}
\Couleur{\forall \Psi \in \mathcal{D},\quad \langle \widetilde{\mathrm{H}} \rangle \equiv  (\mathcal{U}\Psi\ \widetilde{\mid}\ \widetilde{\mathrm{H}} \, (\mathcal{U}\Psi) ) = (\Psi \mid \mathrm{H} \, \Psi )\equiv \langle \mathrm{H} \rangle}.
\ee
Whether $\,\Couleur{\mathrm{H}}\,$ is the Hamiltonian or not, one may prove \cite{A50} that (\ref{H tilde equivt to H - mean}) is true if and only if 
\be \label{H tilde equivt to H}
 \Couleur{\widetilde{\mathrm{H}}=\mathcal{U} \mathrm{H}\mathcal{U}^{-1}}.
\ee
Note that, using the unitarity of $\,\mathcal{U}\,$ (\ref{tilde=isometry}), it is trivial to check that the condition (\ref{H tilde equivt to H}) is {\it sufficient} to ensure the validity of Eq. (\ref{H tilde equivt to H - mean}), i.e., to ensure that all mean values of $\,\Couleur{\mathrm{H}}\,$ and $\,\Couleur{\widetilde{\mathrm{H}}}\,$ are equal. Comparing (\ref{Htilde}) and (\ref{H tilde equivt to H}), we find immediately that the {\it Hamiltonian} operators $\,\Couleur{\mathrm{H}}\,$ and $\,\Couleur{\widetilde{\mathrm{H}}}\,$ are equivalent as operators if and only if 
\be\label{partial_t U=0}
\Couleur{\partial_t \mathcal{U}=0}. 
\ee
Thus: {\it for any wave equation which transforms covariantly under a unitary gauge transformation $\,\mathcal{U}\,$ depending on the time coordinate, the Hamiltonian operator  $\,\mathrm{H}\,$ and its mean values depend on the gauge choice.} Again, note that it is not surprising that by adding a non-zero term $\,-\mathrm{i}\,\mathcal{U}\left[\partial_t \left(\mathcal{U}^{-1}\right)\right]\,$ to $\,\mathcal{U} \mathrm{H}\mathcal{U}^{-1}\,$, one gets different mean values. Indeed, Goldman \cite{Goldman1977} stated without a general proof (but giving an example) that Eq. (\ref{Htilde}) leads to different mean values for $\,\Couleur{\mathrm{H}}\,$ and $\,\Couleur{\widetilde{\mathrm{H}}}\,$ in the time-dependent case. However, it was not obvious that the condition (\ref{H tilde equivt to H}) is indeed {\it necessary} to the equality of all mean values, i.e., to the validity of Eq. (\ref{H tilde equivt to H - mean}). 

\section{Ambiguity of the energy operator}\label{Ambiguity E}

In usual quantum mechanics, the energy operator is simply the Hamiltonian operator H. However, in a time-dependent gravitational field, it happens that H is in general not Hermitian (at least for the covariant Dirac equation \cite{A42,Leclerc2006}). To account for this fact, one defines then the energy operator $\,\mathrm{E}\,$ as the Hermitian part of $\,\mathrm{H}$. It is not hard to guess that the gauge dependence of $\,\mathrm{H}\,$ implies a gauge dependence of $\,\mathrm{E}$, and this is confirmed by a careful examination in the case of the covariant Dirac equation \cite{A43,A50}. Actually, for that equation, one may even prove \cite{A50} that the equality of all mean values of the energy operators  $\,\Couleur{\mathrm{E}}\,$ and $\,\Couleur{\widetilde{\mathrm{E}}}\,$ {\it up to a constant that is independent of the state $\,\Psi \,$} implies $\,\widetilde{\mathrm{E}}=\mathcal{U} \mathrm{E}\mathcal{U}^{-1}\,$ (which is in general not verified when $\,\mathcal{U}\,$ is the unitary transformation associated with a change of the tetrad field). Besides this fact, it was also proved \cite{A43} that the {\it eigenvalue spectrum} of the energy operator depends on the choice of the tetrad field for the covariant Dirac equation. Both of these results are surprising facts.

This is not a small effect: for the covariant Dirac equation, the difference in the mean values of $\,\Couleur{\mathrm{E}} $ for different choices of the tetrad field can be made arbitrarily large \cite{A50}. This is also true with an electromagnetic field \cite{A49}. All of this is already true in a Cartesian chart in a Minkowski spacetime \cite{A50,A49}. It means in particular this \cite{A49}: Unlike the genuine Dirac equation (Dirac's), {\it the covariant Dirac equation cannot predict the energy levels of the hydrogen atom} --- as long as the freedom in the choice of the tetrad field is left too large. In particular, the ambiguity in the energy operator and its mean values is still present, as proved in detail in Ref. \cite{A50}, if one restricts the choice of the tetrad field merely by imposing the ``Schwinger gauge". This is contrary to what had been claimed before by other authors \cite{GorbatenkoNeznamov2011,GorbatenkoNeznamov2013}. 
\footnote{
In brief: in the ``Schwinger gauge", the tetrad field can still be subjected to an arbitrary time-dependent rotation. Around Eq. (35) in Ref. \cite{GorbatenkoNeznamov2014} (whose content is otherwise very close to that of the preprint \cite{GorbatenkoNeznamov2013}), these authors object on my explicit calculation \cite{A50} of the difference in the mean values of E for two Schwinger tetrads that, in their opinion, ``one should also perform averaging over the spin states". But the states relevant to the Dirac equation are four-component states that include the ``spin attribute". The mean value for such a state $\psi $ includes the appropriate ``averaging over the spin states": namely, over those that (in general) are ``mixed" in the state $\psi $  \cite{A51}. 
}

In a very recent paper \cite{Silenko2015}, Silenko also investigates the problem arising from a difference in the mean values of two energy operators (Hermitian Hamiltonians), in a given coordinate system. In his work, the second energy operator is got from the first one by changing the ``representation", i.e., by applying a unitary transformation to the wave function. (That wave function corresponds to some quantum wave equation which {\it a priori} is not necessarily the covariant Dirac equation.) The change of representation that Silenko especially investigates is the one that occurs when going to specifically a Foldy-Wouthuysen representation \cite{FoldyWouthuysen1950}. Thus, the unitary transformation does not arise due to the change of a gauge field. This same problem has been investigated a long time ago by Goldman \cite{Goldman1977} and by Nieto \cite{Nieto1977}. On the other hand, recall that for the covariant Dirac equation, the tetrad field plays the role of a gauge field and its data determines point-dependent Dirac matrices, which do change when one changes that gauge field. Hence, the problem investigated in Refs. \cite{Goldman1977, Silenko2015, Nieto1977} is not the same as the problem investigated in Refs. \cite{A43,A50}. There, we looked to what happens when one changes the point-dependent Dirac matrices (e.g. through the change of the tetrad field) in specifically the covariant Dirac equation. However, it turns out that changing the coefficient fields in the covariant Dirac equation implies that a new representation is got for the Dirac wave function, related to the starting one by a unitary transformation $\,\mathcal{U}\,$: see Eqs. (1)--(5) in Ref. \cite{A50}. Thus, Eqs. (\ref{tilde=isometry}) to (\ref{partial_t U=0})  above do apply to the situations investigated in both groups of works: Refs. \cite{Goldman1977, Silenko2015, Nieto1977} as well as Refs. \cite{A43,A50}.

Equation (31) in Ref. \cite{Silenko2015} gives ``the Dirac Hamiltonian" for a particle in a (possibly nonstationarily) rotating frame [in a spacetime that is flat at least in the domain of the rotating coordinate system, see Eq. (30) there]. To avoid misinterpretations, one should note that this equation gives actually the Dirac Hamiltonian in this rotating frame with a {\it particular choice} of the tetrad field. Namely (in the case of a stationary rotation, $\omega(t) = \mathrm{Constant})$, this is exactly the Hamiltonian got in the rotating frame with ``Ryder's rotating tetrad", Eq. (72) in Ref. \cite{A49}. That Hamiltonian, denoted $\,\mathrm{H}_3\,$ there, has Mashhoon's ``spin-rotation coupling" term. Another choice of tetrad is the ``Cartesian tetrad" (the natural basis of a Cartesian inertial coordinate system, with respect to which the rotating frame is indeed only rotating). Then in the rotating frame one gets the Hamiltonian $\,\mathrm{H}_1\,$ given by Eq. (33) in Ref. \cite{A49}. That one does not have Mashhoon's term and is not physically equivalent to $\,\mathrm{H}_3$, i.e. to the Hamiltonian in Eq. (31) of Ref. \cite{Silenko2015}. (The physical inequivalence of $\,\mathrm{H}_1\,$ and $\,\mathrm{H}_3\,$ has been proved quickly in Ref. \cite{A50}, the paragraph after Eq. (32) there: the proof of inequivalence is indeed the same as the detailed proof given for the case of the Hamiltonians in the inertial frame.) 

To solve the problem of the dependence of the energy mean values on the chosen representation in the non-stationary case, Silenko \cite{Silenko2015} proposes to select the Foldy-Wouthuysen representation as the correct one, essentially because it has a nice semi-classical limit (a point which he demonstrates in detail). He states that the exact definition of a Foldy-Wouthuysen(-type) transformation proposed by Eriksen \cite{Eriksen1958} and further elaborated by Eriksen \& Kolsrud \cite{EriksenKolsrud1960} and de Vries \& Jonker \cite{deVriesJonker1968} can be extended to the nonstationary case. 
\footnote{
Actually, the work of Eriksen \& Kolsrud \cite{EriksenKolsrud1960} also contains an extension to the case with a time-dependent external field, and that extension seems to differ from the one considered in Ref. \cite{Silenko2015}: in the work \cite{EriksenKolsrud1960}, the unitary transformation $U$ leading to the new representation is expressed in terms of an operator $\Lambda $, such that $U\Lambda U^\dagger =\beta $, and in general $\Lambda $ is not equal to the sign operator $\lambda \equiv H.H^{-1/2}$ of the Hamiltonian operator $H$. In the work \cite{Silenko2015}, the transformation $U$ is defined in terms of that sign operator $\lambda $. On the other hand, in the presence of a time-dependent metric and tetrad field (as considered in the latter work), the Hamiltonian does not have directly the form considered in the work \cite{EriksenKolsrud1960}, Eqs. (54)--(55) there. A further discussion could be worth.
}
His proposal \cite{Silenko2015} is at odds (as also noted by him) with the analysis that was made by Goldman \cite{Goldman1977} in the archetypical case of the (genuine) Dirac equation in an inertial frame in a Minkowski spacetime but in a time-dependent electromagnetic field. In the latter case, it seems natural to admit, as did Goldman (albeit with caution), that the correct energy mean values are the ones got from the starting Hamiltonian, which is the (genuine) Dirac Hamiltonian in the electromagnetic field, Eq. (8) in Ref. \cite{Goldman1977}. These mean values differ from those got after the Foldy-Wouthuysen transformation. In our opinion, the fact that in the Foldy-Wouthuysen representation the semi-classical limit is closer to the ``classical" equations of motion (those for a particle with spin yet) than it is in the initial (``Dirac") representation, is not a decisive argument for the choice. Obtaining the classical equations of motion in some limit needs precisely to restrict oneself to some limiting behaviour, which by definition is only at the boundary of the possible behaviours in the quantum domain.  Anyway, let us repeat that the energy ambiguity arising from changing the representation (from the Dirac one to the Foldy-Wouthuysen one) is a different problem than the dependence of the energy mean values on the gauge choice. 

However, we note that in a still recent paper \cite{OST2013}, the energy ambiguity issues due to the presence of a gauge choice (specifically the choice of a tetrad field in the Schwinger gauge) and to a change of the representation (specifically the change to a Foldy-Wouthuysen representation) are simultaneously present. In Ref. \cite{A51}, we have discussed in detail how these inequivalence issues do affect the work \cite{OST2013}. If one would adopt the view \cite{Silenko2015} that the Foldy-Wouthuysen representation is the correct one, then of course he would see no problem in the second inequivalence issue, but the first one would remain.

\section{Energy operator vs. classical energy}\label{Energy}

In this section we summarize a few reasons why the energy operator {\it ought to be} well defined in a given reference frame \cite{A51}, so that it is a real problem that it is not in the presence of a gauge choice. Note first that, even in the time-dependent case, the eigenvalues of $\,\Couleur{\mathrm{E}}\,$ at some (coordinate) time $t$ are the observable values of the energy of the quantum-mechanical system at hand, at this time $t$ \cite{A43}. Moreover, in the stationary case, the eigenvalues of $\,\Couleur{\mathrm{E}}\,$ are associated with stationary solutions of the Schr\"odinger equation (when $\,\Couleur{\mathrm{H}= \mathrm{E}}\,$ is Hermitian). Also, recall that QM has a strong relation to {\it classical Hamiltonian mechanics.} E.g. the standard non-relativistic Schr\"odinger equation is got from the non-relativistic classical Hamiltonian of a test particle:
\be\label{Hamilton-TestParticle}
\Couleur{e=H({\bf p},{\bf x},t)=\frac{{\bf p}^2}{2m}+V({\bf x},t)\equiv T+V,\qquad {\bf p}\equiv m{\bf v}},
\ee
by applying the classical-quantum correspondence
\be\label{Classical-quantum}
\Couleur{e  \rightarrow  +\iC \frac{\partial }{\partial t}, \qquad p_j \rightarrow -\iC \frac{\partial }{\partial x^j}}.
\ee
The correspondence (\ref{Classical-quantum})$_1$ leads also directly to interpret  the Hamiltonian operator $\,\Couleur{\mathrm{H}}\,$ (or its Hermitian part $\,\Couleur{\mathrm{E}}\,$) as the {\it energy operator}. The meaning of the energy operator of QM is hence inherited from classical Hamiltonian mechanics. 

In Newtonian physics and in special relativity, energy is conserved: there is a local energy conservation equation involving the energy density $w$ and its flux $\Mat{\Phi}$,
\be\label{LocalEnergyConservation}
\frac{\partial w}{\partial t}+\mathrm{div}\,\Mat{\Phi }=0.
\ee
Within that continuum with conserved energy, a small piece of matter (its smallness being relative to the problem that is considered) may be modeled as a {\it test particle}. In Newtonian mechanics more specifically, the energy of a test particle is the sum (\ref{Hamilton-TestParticle}) of its kinetic energy and its potential energy in the force field (the latter being assumed to derive from the potential $V$). Clearly, $\,\Couleur{e}\,$ is {\it not} conserved unless $\,\Couleur{V}\,$ is time-independent. Also, $\,\Couleur{e}\,$ depends exactly on the reference frame: e.g., changing the inertial frame changes $\,\Couleur{T}$, but does not change $\,\Couleur{V}$.

Similarly, in a general spacetime in the presence of an electromagnetic potential $\,\Couleur{V_\mu}$, the energy of a test particle is defined as
\be\label{e general}
e\equiv c\breve{p}_0,
\ee
where 
\be\label{def p breve}
\Couleur{\breve{p}^{\mu } \equiv mc \frac{\dd x^\mu }{\dd s}   +  \frac{q}{c}  V^{\mu }}
\ee
and $\,\breve{p}_\mu \equiv g_{\mu \nu} \breve{p}^\nu\, $ [signature $(+---)$], with $\,q\,$ the charge of the particle \cite{A46}. (Also, $\,s\equiv c\tau\, $ where $\,\tau\, $ is the proper time along the trajectory of the particle.) Thus, $\,\breve{p}_{\mu }\,$ being a covector, it follows that $\,\Couleur{e=c\breve{p_0}}\,$ again depends exactly on the reference frame. Moreover, using the dispersion relation
\be \label{GrindEQ__51_}
g^{\mu \nu }  \left( \breve{p}_{\mu }   -  \frac{q}{c}  V_{\mu }  \right) \left( \breve{p}_{\nu }   -  \frac{q}{c}  V_{\nu }  \right) - m^{2} c^{2} =0,
\ee
the energy (\ref{e general}) can be expressed as a function 
\be\label{def H test particle}
H\left({\bf p} , {\bf x}, t \right) \equiv e \equiv  c\breve{p}_0
\ee
with $\,{\bf p}=(p_j)\equiv (-\breve{p}_j)$. This is a Hamiltonian function for the motion of the particle \cite{A46} --- as was already known \cite{Bertschinger1999} in the case without electromagnetic field, i.e., for a free particle in a curved spacetime. Note that in that case there is no potential ($V_\mu =0$), hence the definition (\ref{e general})--(\ref{def p breve}) of the energy is completely unique  for a free test particle in a given reference frame. The Hamiltonian (\ref{def H test particle}) was also already known in the case without gravitational field, i.e., for a charged particle in a Minkowski spacetime \cite{Johns2005}. Clearly, also in the present general case the energy (\ref{e general}) of the test particle is in general not conserved. In conclusion, the fact that the energy operator depends on time in a general spacetime is no justification for it not being unique in a given reference frame at a given time, for the energy of a free test particle also depends on time and it {\it is} unique in a given reference frame at a given time.

 \section{Conclusion}

The energy operator $\,\Couleur{\mathrm{E}}\,$ of a quantum particle is the quantum equivalent of the Hamiltonian energy $\,\Couleur{e}\,$ of a classical test particle. Just as does $\,\Couleur{e}$, it depends precisely on the reference frame, and it is in general not constant. However, in the presence of a gauge choice associated with time-dependent unitary transformations, $\,\Couleur{\mathrm{E}}\,$ is not well defined in a given reference frame. This indicates that the gauge freedom must be reduced to avoid this case. Two proposals along this line have been made for the covariant Dirac equation, each of which provides us with a solution of this non-uniqueness problem \cite{A47,A48}. However, these two solutions are not equivalent. They might be discriminated by the observation or non-observation of the Mashhoon effect \cite{A49}.\\

\noi {\bf Acknowledgement.} I am grateful to Alexander Silenko for a discussion.\\

\vspace{2mm}

\end{document}